\documentclass[prl,twocolumn,showpacs,preprintnumbers,amsmath,amssymb,superscriptaddress]{revtex4}
\usepackage{graphicx}
\usepackage{dcolumn}
\usepackage{bm}
\usepackage{hyperref}

\begin{document}


\title{Theory of single-shot phase contrast imaging in spinor Bose-Einstein condensates}

\author{Ebubechukwu O. Ilo-Okeke}
\affiliation{National Institute of Informatics, 2-1-2
Hitotsubashi, Chiyoda-ku, Tokyo 101-8430, Japan}
\affiliation{Department of Physics, School of Science, Federal University of Technology, P. M. B. 1526, Owerri, Imo State 460001, Nigeria.}

\author{Tim Byrnes}
\affiliation{National Institute of Informatics, 2-1-2
Hitotsubashi, Chiyoda-ku, Tokyo 101-8430, Japan}

\date{\today}

\begin{abstract}
We introduce a theoretical framework for single-shot phase contrast imaging (PCI) measurements of spinor Bose-Einstein condensates.  Our model allows for the simple calculation of the quantum backaction resulting from the measurement, 
and the amount of information that is read out. We find that there is an optimum time $ G\tau \sim 1/N $ for the light-matter interaction ($G $ is the ac Stark shift frequency, $ N $ is the number of particles in the BEC), where the maximum amount of information can be read out from the BEC. A universal information-disturbance tradeoff law $ \epsilon_F \epsilon_G \propto 1/N^2 $ is found where $ \epsilon_F $ is the amount of backaction and $ \epsilon_G $ is the estimation error.  The PCI measurement can also be found to be a direct probe of the quantum fluctuations of the BEC, via the noise of the PCI signal.  
\end{abstract}

\maketitle

Quantum mechanics puts a limit on the amount of information that can be gained by performing a measurement of an unknown state. This concept, arising from the fact that it is impossible to measure an unknown state without disturbing it, has been studied in the context of quantum state estimation \cite{massar95,derka98,paris04}.  Quantum state estimation is typically concerned with how to optimally measure and estimate a given quantum state.  In some contexts it is desirable to limit the amount of backaction the quantum state experiences, and perform an optimal state estimation under this constraint. The other extreme in this continuum is the notion of weak measurements \cite{aharonov88,aharonov07}, where it is in principle possible to perform a measurement without disturbing the system at all, given a large number of copies \cite{aharonov11}.  This information-disturbance tradeoff has been studied in a variety of different classes of quantum states such as single qubits \cite{banaszek01}, multiple copies of qubits \cite{banaszek01b,mista06}, entangled states \cite{sacchi06}, and spin coherent states \cite{sacchi07}.

\begin{figure}
\begin{center}
\includegraphics[width=\columnwidth]{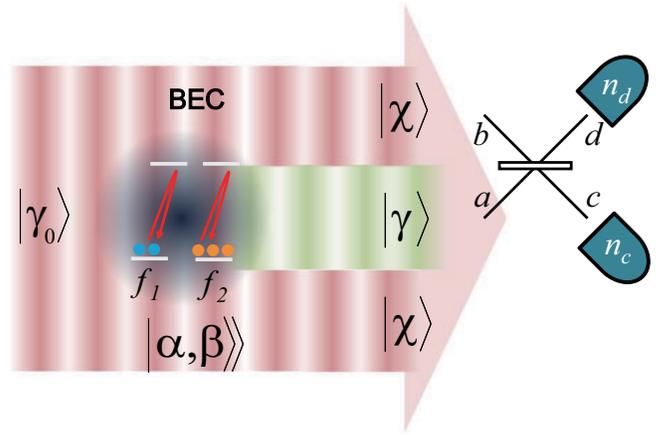}
\caption{\label{fig:non1}(Color online) The phase contrast imaging (PCI) technique.  Coherent light off-resonantly tuned to the excited states of the atoms in the Bose-Einstein condensate (BEC) induces an ac Stark effect entangling the light and the atoms.  The light $\left|\gamma\right>$ passing through the BEC acquires a phase shift, and is interfered with light $\left|\chi\right>$ that does not pass through the BEC.  Homodyne detection on the light beams gives the signal $ n_c - n_d $.  }
\end{center}
\end{figure}

In the context of measurements of atomic Bose-Einstein condensates (BEC), there are several techniques available for probing its state. Commonly, information is read-out via absorption imaging \cite{anderson1995,andrews1997b} or fluorescent imaging~\cite{depue2000} which both result in the destruction of the BEC itself, and are examples of strong or projective measurements. Not all methods result in the destruction of the BEC however, and nondestructive techniques using non-resonant detuned light~\cite{andrews1996,bradley1997b,higbie2005,kohnen2011,brahms2012,gajdacz2013} have been used to measure the properties of ultracold atomic gases~\cite{kohnen2011,brahms2012,gajdacz2013}, as well as small and dense atomic condensates~\cite{andrews1996,bradley1997b,higbie2005} \emph{in-situ}. For instance, in PCI~\cite{andrews1996,bradley1997b,higbie2005} a coherent light illuminates the BEC and a state dependent phase shift develops in the light (see Figure \ref{fig:non1}). By measuring the phase shift via interference, the state of the BEC can be inferred. Because the PCI measurement does not destroy the atomic condensate, it can be applied repeatedly~\cite{andrews1997,meppelink2010} on the same atomic sample.  
Such a technique is a useful resource for information readout in several proposed applications of ultra-cold atomic systems to metrology~\cite{shin2004,wang2005}, quantum information and processing~\cite{cory1997,byrnes2012}. Other methods~\cite{thomsen2002,szigeti2013,hush2013} also use a feedback mechanism to achieve greater control over the backaction induced by the measurement.

In this paper we formulate a theory of PCI measurements at the single-shot level.  We introduce a simple model that allows for the calculation of the backaction on the BEC, and the amount of information that is available by measurement of the 
light. Prior to this paper most works have modeled the effect of the backaction via continuous quantum measurement framework \cite{leonhardt1999,leonhardt2000,dalvit2002,szigeti09,szigeti10}.  These methods deal with the problem by deriving an effective master equation by tracing out the degrees of freedom relating to the measurement device (the light in this case). Other works have discussed fundamental limits on sensitivity \cite{lye03,hope04,hope05}, but do not directly consider the backaction and the available information relating to spinor BECs. We analyze the performance of the PCI measurement with atom-light interaction time and find that there is an optimal time where the maximum amount of information can be extracted from the BECs. Our theory shows that the PCI measurement can be used as a direct probe of the quantum noise of the BEC, and allows for the calculation of the information-disturbance tradeoff.  We find a universal scaling behavior which summarizes the tradeoff behavior, showing that for large $ N $, PCI measurements can asymptotically readout the state of the spinor BEC with negligible backaction.  We also show a direct visualization of the backaction via plotting the Q-function, which shows the extent of the backaction for various interaction times of the PCI measurement. 

We now explain our theoretical model for the PCI measurement.  We assume that the BEC is in a coherent two-component spin coherent state  \cite{arecchi1972,gross12}  of the form
\begin{align}
\left.\left|\alpha,\beta \right>\right> = \frac{1}{\sqrt{N!}} (\alpha f^\dagger_1 + \beta f^\dagger_2)^N \left|0\right>,
\label{spincoherent}
\end{align}
where $ f^\dagger_{1,2} $ are creation operators for atoms in the two states (such as hyperfine ground states), $ N $ is the number of atoms in the BEC, and $ \alpha = \cos (\theta/2) $, $ \beta = e^{i \phi} \sin (\theta/2) $ are Bloch sphere parameters.   
The two-component system can be easily generalized to multi-component BECs and all our results with the exception of Q-function calculations are unchanged. The PCI measurement scheme then proceeds as described in Fig. \ref{fig:non1}. The laser induces an ac Stark shift for each level, which can be described by the Hamiltonian~ \cite{suppl48} 
\begin{align}
H = \hbar G S_z n_a,
\label{acstarkham}
\end{align}
where $ n_a = a^\dagger a $ is the photon number operator and $ S_z = f^\dagger_1 f_1 - f^\dagger_2 f_2 $. The ac Stark shift coupling $ G $ is a second order transition, and can be easily adjusted by either changing the laser amplitude or detuning, thus we consider it to be a free parameter.  Lower order terms in (\ref{acstarkham}) proportional to $ n_a $ have been dropped as they contribute to an irrelevant additional global phase shift of the light. 

The interaction between atoms and light results in entanglement of the light and atomic states
\begin{align}
& e^{-i H \tau/\hbar} \left.\left| \alpha,\beta\right>\right> \left|\gamma_0 \right>  
\nonumber \\
& = \sum_{k=0}^N \sqrt{{N \choose k}} \alpha^k \beta^{N-k} | k \rangle | \gamma e^{i(2k - N) G\tau} \rangle  \left|\chi \right>. 
\label{eq:non2}
\end{align}
Here, $ \left|\gamma_0 \right> $ is the incident coherent state of light of amplitude $ \gamma_0 $, which splits into two components where $\left|\gamma\right> = e^{-|\gamma|^2/2}  e^{\gamma a^\dagger} | 0 \rangle $ is the coherent state of light that passes through the BEC cloud and $ \left|\chi \right> = e^{-|\chi|^2/2}  e^{\chi b^\dagger} | 0 \rangle $ is the light that does not (see Figure \ref{fig:non1}).  $ | k \rangle = \frac{(f_1^\dagger)^k (f_2^\dagger)^{N-k}}{\sqrt{k!(N-k)!}} | 0 \rangle  $ are Fock states of the BEC.  As can be seen from Eq. (\ref{eq:non2}), the interaction between atoms and light causes the light states to undergo a phase rotation. Normally, the ac Stark shift is viewed from the perspective of the shift of the atomic levels due to the light, but here we take the reverse perspective of the atomic influence on the light, and use this to infer information about the atoms. In order to access information about the atoms that is contained in the light states, the phase shift of the light is observed in a homodyne measurement using a 50-50 beam splitter with the transformation $ c = (a - i b)/\sqrt{2}, d = (-ia + b)/\sqrt{2}$. The photons in the modes $c$ and $d$ are then counted giving outcomes $n_c$ and $n_d$ photons respectively.  The resulting unnormalized state is 
\begin{align}
& |\psi_m (n_c, n_d) \rangle  = \frac{e^{\frac{-|\gamma|^2}{2}}}{\sqrt{n_c!}} \frac{e^{\frac{-|\chi|^2}{2}}}{\sqrt{n_d!}} \sum_{k=0}^N \sqrt{{N \choose k}} \alpha^k \beta^{N-k}  \nonumber \\
& \times \left(\gamma e^{i(2k - N) G\tau + i\xi} + i\chi \right)^{n_c} \left(i\gamma e^{i(2k - N) G\tau + i\xi} + \chi \right)^{n_d} | k \rangle ,
\label{postmeasurement}
\end{align}
where $ \xi $ is a phase factor accounting for additional phase shifts between $ a $ and $ b$. We note that our approach differs from existing approaches in that the optical field is treated as a single mode, whereas works such as Ref. \cite{leonhardt1999,leonhardt2000,dalvit2002} treat it as a bath.  In our case the backaction arises due to a projective measurement of the optical field onto number states, while previous works consider 
the illuminating laser to cause the decoherence itself. For a monochromatic laser our single mode approximation is well-justified. 

The signal ${\cal S} $ of the homodyne measurement, equal to the difference between the photon counts on modes $ c $ and $ d $, can be evaluated by taking expectation values with respect to the state (\ref{postmeasurement}).  After some algebra this yields \cite{ilookeke14}
\begin{align}
{\cal S} \equiv  \frac{\left< n_c \right> - \left< n_d \right>}{2|\gamma||\chi|} = e^{-2N| \alpha \beta|^2 G^2\tau^2} \sin( G\tau \langle S^z \rangle_0 + \xi'),    \label{eq:non5}
\end{align}
where we have absorbed constant phase factors into a global offset $\xi' $ and $ \langle S^z \rangle_0 = \langle \langle \alpha, \beta | S^z |  \alpha, \beta \rangle \rangle = N (|\alpha|^2 -  |\beta|^2 ) $ is the $z$-component of the average spin for the state (\ref{spincoherent}). The definition of the signal is normalized such that it is of order unity for typical parameters.  The signal $ {\cal S} $ undergoes oscillations as a function of the average spin of the BEC, as expected.  There is however an unexpected damping term which removes the interference pattern for times $ G\tau> 1/\sqrt{N}$. This occurs because during the entanglement between atoms and light, each photon number state evolves at a different rate and results in accumulation of relative phases between different photon number states. Averaging over the many different number states evolving at different rates gives the exponential decaying amplitude. This suggests that the most suitable time of interaction is $ G\tau \sim 1/N$, a fact we will confirm when examining the information-disturbance tradeoff. 

The variance of the signal may also be evaluated using (\ref{postmeasurement}) to give
\begin{align}
\label{eq:non6}
& (\Delta {\cal S})^2  = 
\frac{|\chi|^2 + |\gamma|^2}{4|\gamma|^2|\chi|^2}  
 + \frac{1}{2}(1 - e^{-8N | \alpha \beta |^2 G^2\tau^2}) \nonumber \\
  & - e^{-4N | \alpha \beta |^2 G^2\tau^2} ( 1- e^{-4N | \alpha \beta |^2 G^2\tau^2} )\sin^2 ( G\tau \langle S^z \rangle_0 + \xi').
\end{align}
The above expression has the simple interpretation that the total variance of the signal is the sum of the fluctuations of the probe beam itself (the first term) and the atomic BEC system. To see this more clearly, let us consider the bright probe beam limit such that the fluctuations of the signal are only due to the atomic BEC.  For times of order $ G \tau \sim 1/N $, we may approximate
\begin{align}
(\Delta {\cal S})^2 \approx \frac{4 | \alpha \beta |^2 \tilde{\tau}^2 }{N } \cos^2 (\langle S^z/N \rangle_0 \tilde{\tau}  + \xi'),
\end{align}
where we introduced a dimensionless time variable $  \tilde{\tau} = G N \tau  $ of order unity $ \tilde{\tau} \sim O(1) $. Considering that the variance of the state (\ref{spincoherent}) is $ (\Delta S^z)^2/N^2 = 4 |\alpha \beta|^2/N $ \cite{byrnes12b} we see that the variance of the PCI signal gives a direct measurement of the quantum fluctuations of the BEC.  The condition for this is that the shot noise of the light is below the noise level of the atoms $ (|\chi|^2 + |\gamma|^2)/4|\gamma|^2|\chi|^2 < \tilde{\tau}^2 /N $.

From the expression derived in (\ref{eq:non5}) and using maximum likelihood estimation theory \cite{dariano2000,paris2004}, we may write an expression for the estimate of the $ S^z $ expectation value $\left<\psi_e\left|S^z/N \right|\psi_e\right>$ as
\begin{align}
\cos \theta_e = \frac{1}{\tilde{\tau}} \arcsin \left[ \frac{(|\gamma|^2 + |\chi|^2)(n_c - n_d)}{2|\chi||\gamma|(n_d + n_c)} \right],
\end{align}
where  $\left|\psi_e\right> = | \cos \frac{\theta_e}{2}  , e^{i\phi_e}\sin \frac{\theta_e}{2} \rangle \rangle $ is the estimated state. $\theta_e$ is the estimated value to the actual parameter $\theta$ of the atomic coherent state from measurement.  As expected, for a PCI measurement in the $ z $-direction (i.e. Eq. (\ref{acstarkham}) involves only the $ S^z $ operator), only information of $ \theta_e $ is obtained.  To obtain an estimate of $ \phi_e $, another PCI measurement in either the $ x $ or $ y $ directions must be made.  We are now in a position to evaluate the information-disturbance tradeoff.  We study the quantities $\epsilon_F$ and $\epsilon_G$ \cite{banaszek2001, sacchi2007} valid for spin coherent states
\begin{equation}
\label{eq:non7}
\begin{split}
\epsilon_F & = \int d\phi\,d\theta\sin\theta \sum_{ {n_c, n_d \atop j=x,y,z} } P_{n_c n_d}  \left[ \frac{\left<\psi_m\left|S^j \right|\psi_m\right>}{N\left<\psi_m| \psi_m \right>}  - \frac{\langle S^j \rangle_0}{N} \right]^2,\\
\epsilon_G & = \int d\phi\,d\theta\sin\theta \sum_{ {n_c, n_d \atop j=x,y,z} }  P_{n_c n_d}  \left[ \frac{\left<\psi_e\left|S^j \right|\psi_e\right>}{N} - \frac{\langle S^j \rangle_0}{N} \right]^2,
\end{split}
\end{equation}
where $P_{n_c n_d} = \langle \psi_m (n_c, n_d)  |\psi_m (n_c, n_d) \rangle $. $\epsilon_F$ gives the disturbance experienced by the amplitude of the coherent state, while $\epsilon_G$ gives the error in the information gained from the measurement of the amplitude of the coherent state. We normalized all spin quantities as $ S^j/N $ such that the distances are defined relative to a unit radius Bloch sphere. Since we are using trace distance, an infidelity measure, a perfect estimate of the state gives $\epsilon_G=0$ and zero disturbance (backaction) corresponds to $\epsilon_F=0$.

\begin{figure}
\begin{center}
\includegraphics[width= 1.1\columnwidth]{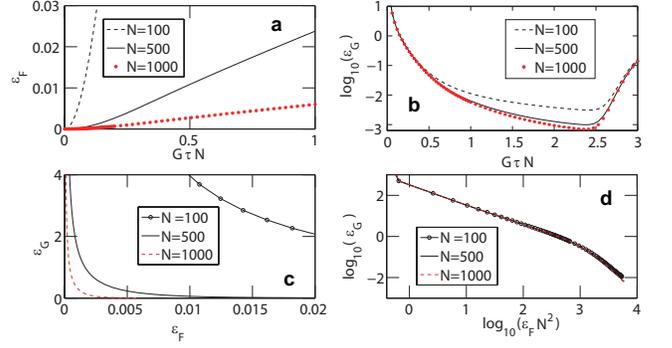}
\caption{\label{fig:non3}(Color online).
(a) The backaction $\epsilon_F$ as a function of the interaction time $ \tilde{\tau} = G \tau N$ and various particle numbers $N $ in the BEC.  (b) The estimation error $\epsilon_G$ as a function of the interaction time.  (c) The information-disturbance tradeoff for various BEC particle numbers as marked. (d) The universal curve showing the behavior (\ref{universal}) of the information-disturbance tradeoff.  Calculations assume parameters $|\chi|^2 = 120$, $|\gamma|^2 = 50 $.  } 
\end{center}
\end{figure}

These two measures are plotted in Fig. \ref{fig:non3} at different values of  total number $N$ of atoms at increasing values of the coupling $G\tau$. In Fig. \ref{fig:non3}(a)  we see that the amount of backaction $\epsilon_F$ generally increases with the interaction time $ \tilde{\tau} $, as expected.  BECs of different particle numbers have different amounts of backaction, with larger systems experiencing less disturbance for the same $ \tilde{\tau}$. This can be understood at the mean field level of the interaction (\ref{acstarkham}), taking the photon operator to be a constant $ n_a = |\gamma|^2 $.  The BEC then experiences a rotation $ e^{-i (|\gamma|^2  \tilde{\tau}/  N) S^z} $, which diminishes with $ N $.  A rotation by an angle $ \phi $ results in a trace distance of 
$ \epsilon_F \approx \phi^2 $ thus we expect that the backaction generally scales as $ \epsilon_F \propto 1/N^2 $. Meanwhile, the error in the estimate $\epsilon_G$ shows the opposite behavior, initially improving with  $ \tilde{\tau} $ (see Figure \ref{fig:non3}(b)).  However, at some point there is no further improvement of the $ \epsilon_G $, which may be attributed to the backaction corrupting the state during the measurement. We find the optimal measurement time to be in the vicinity of 
$ G  \tau N \sim 2.5 $.  The curves have very little dependence on $ N $, which may also be understood at the mean field level.  The light experiences a phase shift $ e^{-i  \langle S^z/N \rangle n_a  \tilde{\tau} } $, which is a fixed rotation as $ \langle S^z/N \rangle $ is always a quantity of order unity.  Thus at fixed $ \tilde{\tau} $, the backaction reduces by a factor $ 1/N^2 $ while the estimation error remains constant. In Figure \ref{fig:non3}(c) we plot the information-disturbance tradeoff for the PCI measurement.  As expected for small disturbance there is a large estimation error and vice versa. As $ N $ is increased, for a given
backaction we may obtain increasingly good estimates of the BEC state.  Due to the above mechanism of the scaling of $\epsilon_{F,G} $ with $ N $, the curves may
be succinctly described by the behavior
\begin{align}
\epsilon_{F} \epsilon_{G} \propto \frac{1}{N^2} .
\label{universal}
\end{align}
To confirm this, we plot in Figure \ref{fig:non3}(d) a log-log plot of the information-disturbance tradeoff, which should appear as a straight line of gradient -1 according to (\ref{universal}).  We find that remarkably all the curves collapse onto the same universal curve, which is a straight line with the expected gradient for several orders of magnitude. The proportionality constant in (\ref{universal}) depends on parameters relating to the measurement $ \gamma, \chi $. In the region with large backaction and small estimation error the line deviates from the linear behavior which we attribute to long evolution times $ G \tau \ge 1/\sqrt{N} $ which has qualitatively different behavior to the main region of interest with $ G \tau \sim 1/N  $.

\begin{figure}
\begin{center}
\includegraphics[width=\columnwidth]{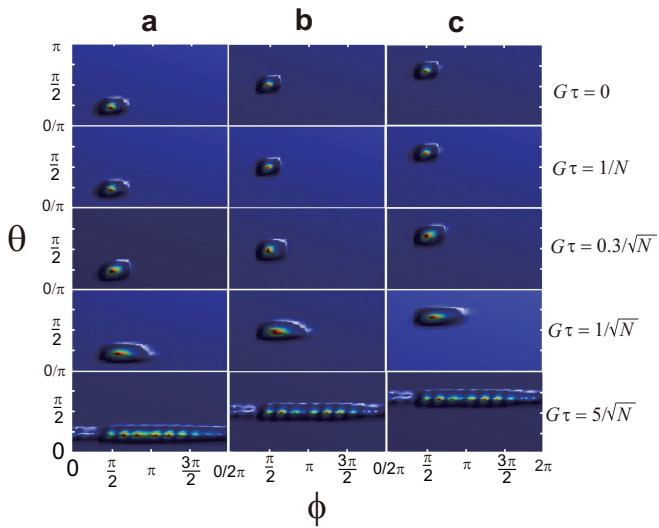}
\caption{\label{fig:non2}(Color online). Q-distribution plots for various initial conditions and light-atom interaction times $G\tau$. The interaction times are as marked and the initial conditions are (a) $\theta = 2\pi/9$, $\phi = \pi/2$; (b) $\theta = \pi/2$, $\phi = \pi/2$; (c) $\theta = 2 \pi/3$, $\phi = \pi/2$.  We use parameters $ N = 300 $, $|\chi|^2 = 80$, $|\gamma|^2 = 4 $.} 
\end{center}
\end{figure}

It is also interesting to examine the backaction in terms of the Q-distribution of the spin coherent states.  This is calculated by averaging over all measurement 
outcomes $n_c, n_d$  in (\ref{postmeasurement}) and using the standard definition of the Q-distribution \cite{gross12} (see Supplementary Information).  The results are presented in Fig~\ref{fig:non2}. 
For interaction times $G\tau \sim 1/N$, the Q-function of the state is hardly affected as depicted in the first three rows of Fig.~\ref{fig:non2}. This is marked by the Q-distribution being Gaussian in this limit with a width that is of the order $\sim 1/\sqrt{N}$. For times approaching $G\tau \sim 1/\sqrt{N}$, the disturbance of the probe is no longer negligible and starts to broaden in the $ \phi $-direction, due to the effect of the entanglement with the light.  Finally for times $G\tau > 1/\sqrt{N} $,  the disturbance is pronounced, with severe broadening such that the state in the $ \phi $ direction is completely distorted with respect to the original state. 
This is reminiscent of the state distribution as found by entangling two spin coherent states \cite{byrnes13}, the difference here being that a further step of measurement has been performed to produce a random mixture of the states in the $ \phi $ direction.  The broadening in the $ \phi $ direction is caused by the ac Stark coupling which is proportional to $ S^z $, the same type of coupling as that considered in Refs. \cite{leonhardt1999,leonhardt2000,dalvit2002}.  Thus our results qualitatively agree with these works, although the precise dynamics will differ. Since the relevant times are $ G \tau \sim 1/N $, the above results imply that in practice one should be able to estimate a given state $ | \alpha, \beta \rangle \rangle $ to extremely good precision, with very little backaction on the state.  This is interesting in terms of quantum information applications since  the content of quantum information stored in the spin coherent states can be determined with good accuracy and very
little disturbance, agreeing with the general results of Ref. \cite{massar95,sacchi07}.

We finally provide some simple parameter estimates to compare our theory to experiment. The magnitude of the ac Stark shift gives $ \delta E = \hbar G \langle n_a \rangle $, where  $ \langle n_a \rangle $ is the average number of photons illuminating the BEC.  The average number of photons is given by $ \langle n_a \rangle = \frac{I A \tau}{h \nu} $, where $ I $ is the intensity of the light, $ A $ is the cross-sectional area of the BEC, and $ \nu $ is the frequency of the light.  This allows for the dimensionless interaction  to be estimated $ G \tau = \frac{2 \pi \delta E \nu}{IA} $, which for the parameters in Ref. \cite{higbie2005} is $ 9 \times 10^{-7} $.  Compared to $ 1/N = 2.5 \times 10^{-7} $ for this experiment, this reproduces the stated result that this experiment is in the minimally-destructive regime.  To avoid the effects of spontaneous decay, we demand that the interaction time be shorter than the effective dephasing time \cite{pyrkov13} $ \tau < \frac{\Delta}{\Gamma G } $, where $ \Delta $ is the detuning of the light and $ \Gamma $ is the spontaneous decay rate -- a condition well satisfied in Ref. \cite{higbie2005}.

In summary, we have presented a simple theory of phase contrast imaging measurement at the single-shot level.  We showed that the optimal light-matter interaction between the BEC and the light occurs for times $ G\tau \sim 1/N $.  
Beyond these times (i.e. $ G\tau \ge 1/\sqrt{N} $) the signal starts to degrade and the state starts to suffer significant backaction,
and no further improvement in the readout of the BEC is obtained.  The information-disturbance tradeoff was calculated and found to satisfy the simple universal relation $ \epsilon_F \epsilon_G \propto 1/N^2 $.  This means that for large $ N $ systems, the PCI measurement should be able to readout the state of the system to quantum limited precision $ \epsilon_G \sim 1/N $, with backaction of the same order $ \epsilon_F \sim 1/N $.  We have also found that the signal can be used as a direct measure of the quantum fluctuations of the system, in the limit of bright probe beams.  The PCI measurement could then potentially be used for quantum information processing applications based on spin coherent states \cite{byrnes2012}, where the state can be readout with very high fidelity while minimally affecting the quantum state, a task which is impossible with standard qubits.  Other possibilities include using the PCI measurement as an explicit realization of a weak measurement by tuning the interaction time $ G \tau $ appropriately.  Although we have not explicitly examined the effects of imperfections due to spontaneous decay \cite{lye03,hope04,hope05}, for typical parameters in the minimally destructive regime this should be a good approximation. We have also assumed that the atoms are in a BEC spin coherent state (\ref{spincoherent}) which is only valid well-below the critical temperature \cite{hush2013}.  Above the critical temperature the atoms are more appropriately described as a spin ensemble, assuming the spin is preserved.  As our coupling only involves the total spin $ S^z $ our arguments should equally hold for such ensembles, as long as the coherence length of the light is larger than the atomic cloud.

\begin{acknowledgments}
This work is supported by the Transdisciplinary Research Integration Center and the Okawa Foundation. 
\end{acknowledgments}


\end{document}